\documentclass[10pt]{article}
\usepackage[top = 2 cm, bottom = 2.5 cm, left = 2.5 cm, right = 2 cm]{geometry}
\usepackage{authblk, cite, color, amssymb, amsmath, enumitem}
\usepackage[hypertex, colorlinks = true, linkcolor = blue, citecolor = red]{hyperref}
\usepackage[dvipsnames]{xcolor}

\begin{document}

\title{The bimetric model with an informational metric tensor}

\author[1,2]{Merab Gogberashvili \thanks{gogber@gmail.com}}
\affil[1]{Javakhishvili State University, 3 Chavchavadze Ave., Tbilisi 0179, Georgia}
\affil[2]{Andronikashvili Institute of Physics, 6 Tamarashvili St., Tbilisi 0177, Georgia}
\maketitle

\begin{abstract}

We consider a hybrid bimetric model where, in addition to the ordinary metric tensor that determines geometry, an informational metric is introduced to describe the reference frame of an observer. We note that the local information metric being Minkowskian explains one of the key aspects of the Einstein equivalence principle. Our approach has the potential to justify the three-dimensional nature of physical space and address the gravitational energy puzzle. Furthermore, it appears to be free of ghost instabilities in the matter sector, as the second metric tensor couples exclusively to the observer and is non-dynamical.

\vskip 2mm
\noindent
PACS numbers: 04.50.Kd (Modified theories of gravity); 02.40.Ky (Riemannian geometries); 89.70.+c (Information theory and communication theory)

\vskip 1mm
\noindent
Keywords: Bimetric model; Gravitational energy; Entropy
\end{abstract}


General Relativity has been successful in describing the gravity of stars and planets \cite{Will:2014kxa}, but it faces challenges when applied to different scales and encounters theoretical problems related to local energy-momentum conservation. These issues have motivated researchers to explore alternative theories of gravity. One such alternative is bimetric theories \cite{Hassan:2011zd}, which involve two fundamental metric tensors at each point in spacetime. This allows bimetric models to address the gravitational energy problem, since under a change of coordinates Christoffel symbols behave as pseudo-tensors, while the difference of connections constructed with different metric tensors is a tensor. Bimetric theories also have the potential to offer insights into various challenges and questions in gravitational physics, including issues related dark matter \cite{Aoki:2014cla, Babichev:2016hir, Kolb:2023dzp, Gialamas:2023aim}, black holes \cite{Volkov:2012wp}, and cosmological phenomena \cite{Gialamas:2023lxj} (see the reviews \cite{Schmidt-May:2015vnx, Clifton:2011jh} and references therein).

Bimetric theories also have the potential to deal with some long-range problems, e.g. with dark matter \cite{Aoki:2014cla, Babichev:2016hir, Kolb:2023dzp, Gialamas:2023aim}, black holes \cite{Volkov:2012wp}, inflation \cite{Gialamas:2023lxj} (see the reviews \cite{Schmidt-May:2015vnx, Clifton:2011jh} and references therein). However, many models with two interacting second-rank tensors encounter problems with ghost instabilities. Additionally, Lovelock's theorem severely restricts possible bimetric extensions \cite{Koyama:2015vza}.

Note that standard differential geometry already considers two kinds of geometrical quantities at any point, as it distinguishes between a manifold and a coordinate system, as well as between a manifold and frame objects. A manifold is seen as a set of points where it is possible to embed coordinate systems of an observer. A clear distinction between active and passive diffeomorphisms is also formulated \cite{Sundermeyer:2014kha, Rovelli:2004tv}. Under passive transformations, an observer changes the view (coordinate system), but the physical world (represented, for example, by tensors) being observed does not change. In contrast, an active transformation transforms tensors while keeping the observer fixed. The principle of active general covariance states that the dynamics of a manifold, where geometrical structures can be expressed in various ways, is equivalent to the passive view, where structures can be described using different coordinate systems. It is assumed that the observer's information does not impact the frame, and the active and passive views are kinematically equivalent. This assumption leads to the concept of matter objectivity, which emphasizes frame-indifference (see the review \cite{MFI}). However, according to General Relativity, the notion of an unchanging spacetime manifold during a coordinate transformation can only be understood as an idealization, assuming that the change of the frame is not coupled to a massive observer. In reality, the distribution of mass determines the geometry of spacetime, and conversely, the structure of the spacetime manifold determines the mass distribution within it.

Information metrics also are indirectly incorporated in the fabric of standard General Relativity to resolve, for example, the problems raised by the Einstein hole argument and to remove spurious solutions \cite{hole}. By introducing an information about metric, these problems can be resolved, preventing the existence of multiple solutions due to gauge invariance. In the process of solving Einstein's equations, an observer constructs a template metric that defines the geometric properties of coordinates and largely determines their physical interpretation. The final step is to substitute the template metric into the Einstein equations to determine the remaining parameters and ensure consistency with the boundary conditions defined by the observer \cite{Macdonald}.

The statement that geometry is objective regardless of the way it is observed is not universally true. Examples are non-associative geometries \cite{Gogberashvili:2022kzw}, the Unruh effect and geodesics close to horizons \cite{Sus-Lin}. Typically, the choice of the reference system and local coordinates is arbitrary since none of the metrics has preferred status. However, for extended systems and metrics that lead to geometrical singularities, a transition to another coordinate system is generally unacceptable.

Based on ideas traced back to Poincaré \cite{Poincare}, the geometry of spacetime depends on the properties of measuring instruments (frame of reference). It is expected that there exists a relationship between the metric of spacetime and the reference frame used. Physical quantities need to be compared with experimental observations, and our models must be based on an observer-dependent formulation. According to Landauer's and Brillouin's principles, the abstract concepts of observer information can be associated with energy or mass \cite{Info, Ilgin:2014dua}, which can influence spacetime geometry. Any measurement, including a thought experiment, leads to an increase in energy. While informational distortions of the metric are negligible for small systems, they can have significant effects on massive objects (such as black holes) \cite{Gogberashvili:2021gfh} and cosmological scales \cite{Gogberashvili:2022ttz}. Let's consider coordinates where the Christoffel symbols are zero and energy is conserved. In other coordinate systems, we may only have covariant conservation (i.e., non-conservation), but following the spirit of Landauer's principle we can restore the conservation of energy by introducing information energy corresponding to these coordinate transformations.

Understanding the importance of the concept of observer to describe even classical geometry, in this paper we propose to consider a hybrid bimetric model. In addition to the ordinary metric tensor that directly couples to matter and determines the geometry we probe, we suggest to introduce an informational metric that describes the reference frame of an observer \cite{Gogberashvili:2016wsa, Gogberashvili:2022cam, Amari-1, Amari-2, Rao, Caticha}. This approach can solve the gravitational energy puzzle, similar to standard bimetric models, and also appearing to be free of ghost instabilities in the matter sector, since the second metric tensor couples only to the observer and is non-dynamical.


To describe any physical system, it is necessary to define an observer's frame and local geometry. In the information approach, the key concept is the distinguishability of matter particles, which can be quantified by probability distributions $p(X^i)$ of some information coordinates $X^i$ that characterize particles. The corresponding entropy given by
\begin{equation} \label{S}
{\cal S} = - \ln ~p(X^i)~.
\end{equation}
The information distance between neighboring distributions, $p(X^i)$ and $p(X^i + dX^i)$, can be described by the dimensionless Fisher-Rao metric \cite{Rao, Amari-1, Amari-2, Caticha},
\begin{equation} \label{Info-metric}
d\mathfrak{L}^2 = g_{nm} dX^ndX^m ~.
\end{equation}
A small value of the interval $\mathfrak{L}$ indicates that particles with $X^i$ and $X^i + dX^i$ are difficult to distinguish. Thus, the concept of distance, including the metric $g_{nm}$ as its infinitesimal version, can be replaced by the notion of observability. Smaller relative entropy between the distributions of two particles implies a shorter distance between them.

It is worth noting that the informational parameters $X^i$ are arbitrary since the distributions of particles can be relabeled. Under these changes, the information metric $g_{nm}$ in (\ref{Info-metric}), which is done by
\begin{equation} \label{g=dS}
g_{nm} = \frac {\partial^2 {\cal S}(X^i)}{\partial X^n \partial X^m}~,
\end{equation}
transforms as a second-rank tensor \cite{Rao, Amari-1, Amari-2, Caticha}. The coefficients of connection for (\ref{g=dS}) have the simple form:
\begin{equation} \label{Gamma}
{\Gamma^j}_{nm} = \frac 12 \, \frac {\partial^3 {\cal S}(X^i)}{\partial X_j \partial X^n \partial X^m}~.
\end{equation}
Defined by \eqref{g=dS} informational space is Riemannian \cite{Rao, Amari-1, Amari-2, Caticha} and equipped with a connection \eqref{Gamma} which is metric compatible,
\begin{equation}
\nabla_i g_{mn} = \frac {\partial g_{mn}}{\partial x^i} - {\Gamma^j}_{i m} g_{j n} - {\Gamma^j}_{i n}g_{m j} = 0
\end{equation}
($\nabla_i$ denotes the covariant derivative), and torsion-free,
\begin{equation}
{\Gamma^i}_{nm} = {\Gamma^i}_{mn}~.
\end{equation}
Using the definitions \eqref{g=dS} and \eqref{Gamma}, one can check that the information space (\ref{Info-metric}) is also flat,
\begin{equation}
R_{ij nm} = 0~,
\end{equation}
implying that the metric tensor (\ref{g=dS}) can always be written in the Euclidean form:
\begin{equation} \label{info-local}
g_{ij} = \delta_{mn}\frac {\partial \bar X^m}{\partial X^i} \frac{\partial \bar X^n}{\partial X^j} ~,
\end{equation}
where $\bar X^m$ represents new coordinates and $\delta_{mn}$ is the Kronecker delta.

We note that the maximal entropy of a random variable with $n$ realizations is typically ${\cal S} \sim \ln n$. However, in the case of an ensemble of $N$ entangled identical particles, it appears that the entropy scales as ${\cal S} \sim \ln N^3$ \cite{Gogberashvili:2007dw, Gogberashvili:2010em, Gogberashvili:2013cea, Gogberashvili:2014ora}. This suggests that when determining information distances using probability distributions $p(X^i)$ of information variables $X^i$, an observer obtains three copies of the entropy $\sim \ln N$, and requires three information coordinates $X^i$ ($i = 1, 2, 3$). Each of the three information coordinates captures an independent aspect of the information associated with entangled particles, and this may be correlated with the three-dimensional nature of physical space. This concept resonates with certain aspects of knot theory  \cite{knots}, which also relies on three dimensions to form stable structures capable of retaining information \cite{Berera:2015yna}.

The Riemannian information 3-space (\ref{Info-metric}) does not capture the light-cone structure. However, entropy is also related to time and the transfer of information. There should be a limit to how fast information can propagate through the finite Universe. Let's consider a physical system within a spherical volume of radius $r$ that contains a total entropy ${\cal S}$. Neglecting entropy generated by internal sources, we can express the rate of change of entropy with respect to time as \cite{Hartman:2013qma, Liu:2013iza, Liu:2013qca}:
\begin{equation} \label{dot S}
\frac{d{\cal S}}{dt} = \frac{3}{r}\, v \,{\cal S}_m~,
\end{equation}
where ${\cal S}_m$ denotes the flux of matter entropy through the boundary, and $3/r$ accounts for the surface-to-volume ratio. In this equation, a quantity $v$ with dimensions of velocity appears, which governs the rate of information transfer. To estimate the value of the information speed $v$ for the Universe, we can apply (\ref{dot S}) to the Hubble volume. By using the entropy expression resembling that of a black hole, the Euler thermodynamic equation for cosmology, and the de Sitter temperature associated with the horizon, the relation (\ref{dot S}) for the Hubble sphere reduces to the standard Raychaudhuri or the second Friedmann equation \cite{Gogberashvili:2022cam}, but only if $v$ is identified with the speed of light $c$. Therefore, within the information formalism, the speed of light $c$ can be interpreted as a fundamental parameter that characterizes the velocity of information transfer in the Universe. This interpretation aligns with the established understanding in physics, where $c$ is regarded as the ultimate speed limit, defining the maximum rate at which information can propagate through space.

The relationship between entropy and radius of a spherical volume, as expressed by the Bekenstein-Hawking formula ${\cal S} \sim r^2$, provides a natural connection between the notion of distance and entropy. By relating the dimensionless information distance $\mathfrak{L}$ to the intervals in Euclidean 3-space and introducing the information speed $c$, it is possible to embed (\ref{Info-metric}) into a 4-dimensional Minkowski spacetime \cite{Gogberashvili:2022cam}. Since the information space (\ref{info-local}) is Euclidean, it implies that the local geometry perceived by an observer appears to be Minkowskian and naturally exhibits a (1+3)-splitting. Consequently, the concept of local relativistic invariance in an observer's frame, which is one of the key ingredients of the Einstein equivalence principle, follows naturally from the information formalism.

Mathematically, an observer can model four information coordinates using number lines. It is known that the real line is a one dimensional Euclidean vector space over the field of real numbers (i.e. over itself). At the observer's point, which is associated with physical frames (to elementary particles in infinitesimal cases), four linearly independent unit vectors $e_a$ ($a=0,1,2,3$) can be introduced. Local distances (or distinguishabilities) are then defined by the Minkowski metric tensor $\eta_{ab}$, for which the Riemann tensor is identically zero. To describe the surrounding matter, the observer needs to consider geometric metric tensors $g_{\mu\nu}(x^\alpha)$ ($\mu,\nu=0,1,2,3$) that are solutions of the classical Einstein equation. In this bimetric model, the observer considers two metric tensors: $\eta_{ab}(x^c)$ for the local Minkowskian information space and $g_{\mu\nu}(x^\alpha)$ for the general geometry of spacetime. The interplay between the two Riemannian metric tensors allows the observer to incorporate both the local information coordinates and the gravitational effects of the surrounding matter.

To relate the information basis to the coordinate basis, transition coefficients (often referred to as tetrads) $h^\mu_a$ are introduced:
\begin{equation}
x^a = h_\mu^a x^\mu, \quad e_a = h^\mu_a e_\mu,
\end{equation}
where $x^a$ and $x^\mu$ represent the local and geometric coordinates, respectively, and $e_\mu$ are orthonormal geometric basis vectors. These basis vectors can undergo parallel transitions to neighboring points of observation or to different observers. In order to incorporate the information tetrads $h^\mu_a$ into the parallel transport of vectors in curved spacetime with the metric tensor $g_{\mu\nu}$, we aim to construct a generalized version of covariant derivatives that accounts for the local information indices. By employing the generalized covariant derivative, the observer can account for the local information structure while describing the curvature of spacetime and the parallel transport of tensors.

We start by considering the covariant derivative of the Minkowskian basis vector $e_a$ along the curved spacetime basis vector $e_\mu$. The derivative of a vector cannot be simply the ordinary derivative because any shift in the vector corresponds to a change in the information of an observer or a change of the observer itself. Therefore, we need to define a derivative that takes this into account. We suppose that the desired information derivative is analogous to a covariant derivative and is a linear operator satisfying the Leibniz rule. The rate at which $e_a$ changes when it is moved along the geometrical basis vector (covariant derivative) should be a vector with respect to another observational point and can be expressed as a linear combination of the initial vector $e_a$.

In our bimetric model, the basis vector of the local Minkowskian frame $e_a$, after being parallel transported along the geometrical basis vector $e_\mu$ to another observer's frame, changes due to two factors: the effect of Riemannian curvature and the relabeling of information coordinates in local Minkowskian spacetimes. Therefore, we can write the covariant derivative as:
\begin{equation}
\nabla_\mu e_a = h^b_\mu \widetilde{\nabla}_b e_a + h^\nu_a \nabla_\mu e_\nu = \mathbf{\Gamma}^c_{\mu a} e_c~,
\end{equation}
where $ \widetilde{\nabla}_b$ denotes the covariant derivative in local Minkowski spacetime, $\nabla_\mu$ represents the ordinary Riemannian covariant derivative, and $\mathbf{\Gamma}^c_{\mu a}$ are the components of the Levi-Civita connection (Christoffel symbols). From this relation, we can determine the expression for the geometric coordinate derivatives:
\begin{equation}
\nabla_\mu e_\nu = \Gamma^c_{\mu\nu} e_c = \left( \mathbf{\Gamma}^\alpha_{\mu\nu} - \big\{ ^{\,\alpha}_{\mu\nu} \big\}\right) e_\alpha~,
\end{equation}
where $\big\{ ^{\,\alpha}_{\mu\nu} \big\}$ denotes the connection coefficients of Minkowski spacetime. It is worth noting that the connection coefficients for $e_\nu$ and $e_a$ have different signs, and the covariant derivative for a vector in any chosen reference frame should include the difference between two Christoffel symbols: one corresponding to the vector itself and another constructed using the information metric of the observer defining the reference frame,
\begin{equation}
\Gamma^\alpha_{\mu\nu} = \mathbf{\Gamma}^\alpha_{\mu\nu} - \big\{^{\,\alpha}_{\mu\nu} \big\}~.
\end{equation}
This formulation allows for the consistent treatment of the covariant derivative and captures the interplay between the Minkowskian information space and the curved spacetime geometry.

In order to properly describe a physical system, it is necessary to choose an appropriate observer's frame. Since information coordinates are modeled by one-dimensional Euclidean vector spaces, the natural choice is the frame 4-vector. We can extend the expression for the evolution of entropy (\ref{dot S}) to its relativistic generalization by introducing the entropy variation 4-vector:
\begin{equation}
s_\nu = \frac{\partial {\cal S}}{\partial x^\nu}~.
\end{equation}
This 4-vector characterizes any classical reference system and observer, and its components represent the rates of change of entropy with respect to the spacetime coordinates. In the context of relativity, to each reference frame, we can associate an intrinsic, covariantly conserved vector field \cite{Aoki:2020prb}:
\begin{equation}
u^\nu = T^\nu_\mu s^\mu~, \qquad T^\nu_\mu \partial_\nu s^\mu = 0~,
\end{equation}
where $T^\nu_\mu$ is the energy-momentum tensor of the system. This vector field represents the flow of entropy and energy-momentum associated with the reference frame. By constructing such an intrinsic vector field associated with a reference frame, we establish a covariant formulation that incorporates the evolution of entropy and its relation to the energy-momentum distribution within the system.

In this model, the universe is described as consisting of two components: ordinary matter and an observer's information field $s^\nu$. The geometry of spacetime is understood in relational terms, where informational quantities serve as templates for the coordinates of matter fields, and the informationally defined metric determines how the observer locally experiences spacetime. This approach has some analogies with the Einstein-aether theory \cite{Eling:2004dk}, where a spacetime is endowed with a metric and a timelike vector field (the aether) that defines a preferred reference frame and, hence, violates Lorentz invariance. In contrast, in our model the preferred frame is not fundamental but emerges from the observer's information field.


In summary, the bimetric model considered in this paper, incorporates an additional informational metric tensor alongside the conventional one, which governs spacetime geometry. This informational metric is locally Minkowskian by definition and explains validity of the Einstein equivalence principle. By considering both metrics, our approach holds promise in resolving the enigma surrounding gravitational energy, akin to other bimetric frameworks. Importantly, the coupling of the second metric tensor to the observer only, without dynamical properties, suggests that our model is devoid of ghost instabilities in the matter sector.

Overall, the introduction of informational metrics and the incorporation of information theory into gravity are intriguing concepts that warrant further exploration. These ideas may lead to novel insights into the nature of spacetime and its relationship with observers and information. However, as with any new theoretical framework, rigorous testing and validation against empirical observations are essential steps in assessing its viability as a description of the physical universe.


\section*{Declarations:}

\subsection*{Ethical Approval}
Not applicable.

\subsection*{Conflict of interest}
M.G. declares no conflicts of interest.

\subsection*{Authors' contributions}
M.G. wrote the paper.

\subsection*{Funding}
Not applicable.

\subsection*{Availability of data and materials}
There are no data associated with this article.



\begin{thebibliography}{99}

\bibitem{Will:2014kxa} C.~M.~Will,
``The Confrontation between General Relativity and experiment,''
Living Rev. Rel. \textbf{17} (2014) 4,
doi: 10.12942/lrr-2014-4
[arXiv: 1403.7377 [gr-qc]].

\bibitem{Hassan:2011zd} S.~F.~Hassan and R.~A.~Rosen,
``Bimetric gravity from ghost-free massive gravity,''
JHEP \textbf{02} (2012) 126,
doi: 10.1007/JHEP02(2012)126
[arXiv: 1109.3515 [hep-th]].

\bibitem{Aoki:2014cla} K.~Aoki and K.~i.~Maeda,
``Dark matter in ghost-free bigravity theory: From a galaxy scale to the universe,''
Phys. Rev. D \textbf{90} (2014) 124089,
doi: 10.1103/PhysRevD.90.124089
[arXiv: 1409.0202 [gr-qc]].

\bibitem{Babichev:2016hir} E.~Babichev,  {\it et al.}
``Bigravitational origin of dark matter,''
Phys. Rev. D \textbf{94} (2016) 084055,
doi: 10.1103/PhysRevD.94.084055
[arXiv: 1604.08564 [hep-ph]].

\bibitem{Kolb:2023dzp} E.~W.~Kolb, S.~Ling, A.~J.~Long and R.~A.~Rosen,
``Cosmological gravitational particle production of massive spin-2 particles,''
JHEP \textbf{05} (2023) 181,
doi: 10.1007/JHEP05(2023)181
[arXiv: 2302.04390 [astro-ph.CO]].

\bibitem{Gialamas:2023aim} I.~D.~Gialamas and K.~Tamvakis,
``Bimetric-affine quadratic gravity,''
Phys. Rev. D \textbf{107} (2023) 104012,
doi: 10.1103/PhysRevD.107.104012
[arXiv: 2303.11353 [gr-qc]].

\bibitem{Volkov:2012wp} M.~S.~Volkov,
``Hairy black holes in the ghost-free bigravity theory,''
Phys. Rev. D \textbf{85} (2012) 124043,
doi: 10.1103/PhysRevD.85.124043
[arXiv: 1202.6682 [hep-th]].

\bibitem{Gialamas:2023lxj} I.~D.~Gialamas and K.~Tamvakis,
``On the bimetric Starobinsky model,''
[arXiv: 2307.05673 [gr-qc]].

\bibitem{Schmidt-May:2015vnx} A.~Schmidt-May and M.~von Strauss,
``Recent developments in bimetric theory,''
J. Phys. A \textbf{49} (2016) 183001,
doi: 10.1088/1751-8113/49/18/183001
[arXiv: 1512.00021 [hep-th]].

\bibitem{Clifton:2011jh} T.~Clifton, P.~G.~Ferreira, A.~Padilla and C.~Skordis,
``Modified gravity and cosmology,''
Phys. Rept. \textbf{513} (2012) 1,
doi: 10.1016/j.physrep.2012.01.001
[arXiv: 1106.2476 [astro-ph.CO]].

\bibitem{Koyama:2015vza} K.~Koyama,
``Cosmological tests of modified gravity,''
Rept. Prog. Phys. \textbf{79} (2016) 046902,
doi: 10.1088/0034-4885/79/4/046902
[arXiv: 1504.04623 [astro-ph.CO]].

\bibitem{Sundermeyer:2014kha} K.~Sundermeyer,
                {\it Symmetries in Fundamental Physics},
                Fundamental Theories of Physics, Vol. 176
                (Springer, NY 2014)
                doi: 10.1007/978-94-007-7642-5.

\bibitem{Rovelli:2004tv} C.~Rovelli,
                        {\it Quantum Gravity},
                        Cambridge Monographs on Mathematical Physics
                        (Cambridge University Press, NY 2004)
                        doi: 10.1017/CBO9780511755804.

\bibitem{MFI} M.~Frewer,
"More clarity on the concept of material frame-indifference in classical continuum mechanics,"
Acta mechanica \textbf{202} (2009) 213,
doi: 10.1007/s00707-008-0028-4.

\bibitem{hole} R.~Torretti,
              {\it Relativity and Geometry}
              (Dover, NY 1996).

\bibitem{Macdonald} A.~Macdonald,
``Einstein’s hole argument,''
 Am. J. Phys. \textbf{69} (2001) 223,
 doi: 10.1119/1.1308265.

\bibitem{Gogberashvili:2022kzw} M.~Gogberashvili,
``Algebraical entropy and arrow of time,''
Entropy \textbf{24} (2022) 1522,
doi: 10.3390/e24111522
[arXiv: 2211.00501 [physics.gen-ph]].

\bibitem{Sus-Lin} L.~Susskind and J.~Lindesay,
                 {\it An Introduction to black holes, information, and the string theory revolution: the holographic universe}
                 (World Scientific, Singapore 2005).

\bibitem{Poincare} H.~Poincar\'{e},
                  {\it Derni\`{e}res pens\'{e}es}
                  (Flammarion, Paris 1913).

\bibitem{Info} L.~Herrera,
``Landauer principle and General Relativity,''
Entropy \textbf{22} (2020) 340,
doi: 10.3390/e22030340
[arXiv: 2003.07436 [gr-qc]].

\bibitem{Ilgin:2014dua} I.~Ilgin and I.~S.~Yang,
``Energy carries information,''
Int. J. Mod. Phys. A \textbf{29} (2014) 1450115,
doi: 10.1142/S0217751X14501152
[arXiv: 1402.0878 [hep-th]].

\bibitem{Gogberashvili:2021gfh} M.~Gogberashvili and B.~Modrekiladze,
``Probing the Information-Probabilistic Description,''
Int. J. Theor. Phys. \textbf{61} (2022) 149,
doi: 10.1007/s10773-022-05129-3
[arXiv: 2105.05034 [gr-qc]]

\bibitem{Gogberashvili:2022ttz} M.~Gogberashvili,
``Fixing cosmological constant on the event horizon,''
Eur. Phys. J. C \textbf{82} (2022) 1049,
doi: 10.1140/epjc/s10052-022-11033-1
[arXiv: 2301.04334 [gr-qc]].

\bibitem{Gogberashvili:2016wsa} M.~Gogberashvili,
``Information-probabilistic description of the universe,''
Int. J. Theor. Phys. \textbf{55} (2016) 4185,
doi: 10.1007/s10773-016-3045-4
[arXiv: 1504.06183 [physics.gen-ph]].

\bibitem{Gogberashvili:2022cam} M.~Gogberashvili,
``Towards an information description of space-time,''
Found. Phys. \textbf{52} (2022) 74,
doi: 10.1007/s10701-022-00594-6
[arXiv: 2208.13738 [physics.gen-ph]].

\bibitem{Rao} C.~R.~Rao,
``Information and the accuracy attainable in the estimation of statistical parameters,''
Bull. Calcutta Math. Soc. \textbf{37} (1945) 81;
             In: {\it Breakthroughs in Statistics: Foundations and basic theory},
             S.~Kotz and N.~L.~Johnson (eds), Springer Series in Statistics, pp. 235-247
             (Springer, NY 1992)
             doi: 10.1007/978-1-4612-0919-5\_16.

\bibitem{Amari-1} Sh.-I.~Amari,
               {\it Differential-Geometrical Methods in Statistics},
               Lecture Notes in Statistics, Vol. 28
               (Springer-Verlag, Berlin Heidelberg 1985).

\bibitem{Amari-2} Sh.-I.~Amari and H.~Nagaoka,
                {\it Methods of Information Geometry},
                Translations of Mathematical Monographs, Vol. 191
                (American Math. Soc., 2000).

\bibitem{Caticha} A.~Caticha,
"Lectures on probability, entropy and statistical physics,"
[arXiv: 0808.0012 [physics.data-an]].

\bibitem{Gogberashvili:2007dw} M.~Gogberashvili,
``Machian solution of hierarchy problem,''
Eur. Phys. J. C \textbf{54} (2008) 671,
doi: 10.1140/epjc/s10052-008-0559-9
[arXiv: 0707.4308 [hep-th]].

\bibitem{Gogberashvili:2010em} M.~Gogberashvili and I.~Kanatchikov,
``Machian origin of the entropic gravity and cosmic acceleration,''
Int. J. Theor. Phys. \textbf{51} (2012) 985,
doi: 10.1007/s10773-011-0971-z
[arXiv: 1012.5914 [physics.gen-ph]].

\bibitem{Gogberashvili:2013cea} M.~Gogberashvili,
``On the dynamics of the ensemble of particles in the thermodynamic model of gravity,''
J. Mod. Phys. \textbf{5} (2014) 1945,
doi: 10.4236/jmp.2014.517189
[arXiv: 1309.0376 [gr-qc]].

\bibitem{Gogberashvili:2014ora} M.~Gogberashvili and I.~Kanatchikov,
``Cosmological parameters from the thermodynamic model of gravity,''
Int. J. Theor. Phys. \textbf{53} (2014) 1779,
doi: 10.1007/s10773-013-1976-6
[arXiv: 1210.4618 [physics.gen-ph]].

\bibitem{knots} C.~C.~Adams,
               {\it The Knot Book: An Elementary Introduction to the Mathematical Theory of Knots}
               (American Mathematical Society, 2004).

\bibitem{Berera:2015yna} A.~Berera, {\it et al.}
``Knotty inflation and the dimensionality of spacetime,''
Eur. Phys. J. C \textbf{77} (2017) 682,
doi: 10.1140/epjc/s10052-017-5253-3
[arXiv: 1508.01458 [hep-ph]].

\bibitem{Hartman:2013qma} T.~Hartman and J.~Maldacena,
``Time evolution of entanglement entropy from black hole interiors,''
JHEP \textbf{05} (2013) 014,
doi: 10.1007/JHEP05(2013)014
[arXiv: 1303.1080 [hep-th]].

\bibitem{Liu:2013iza} H.~Liu and S.~J.~Suh,
``Entanglement tsunami: Universal scaling in holographic thermalization,''
Phys. Rev. Lett. \textbf{112} (2014) 011601,
doi: 10.1103/PhysRevLett.112.011601
[arXiv: 1305.7244 [hep-th]].

\bibitem{Liu:2013qca} H.~Liu and S.~J.~Suh,
``Entanglement growth during thermalization in holographic systems,''
Phys. Rev. D \textbf{89} (2014) 066012,
doi: 10.1103/PhysRevD.89.066012
[arXiv: 1311.1200 [hep-th]].

\bibitem{Aoki:2020prb} S.~Aoki, T.~Onogi and S.~Yokoyama,
``Conserved charges in general relativity,''
Int. J. Mod. Phys. A \textbf{36} (2021) 2150098,
doi: 10.1142/S0217751X21500986
[arXiv: 2005.13233 [gr-qc]].

\bibitem{Eling:2004dk} C.~Eling, T.~Jacobson and D.~Mattingly,
``Einstein-Aether theory,''
in {\it Deserfest: A Celebration of the Life and Works of Stanley Deser}, p. 163
(World Scientific, Singapore 2006)
[arXiv: gr-qc/0410001 [gr-qc]].

\end{thebibliography}
\end{document}